\documentclass[manuscript,12pt]{aastex}
\newcommand{\nature}{{\it Nature\ }}
\newcommand{\apjn}{{\it Astrophys. J.\ }}
\newcommand{\apjsn}{{\it Astrophys. J. Supp.\ }}
\newcommand{\aan}{{\it Astron. Astrophys.\ }}
\newcommand{\ajn}{{\it Astron. J.\ }}

\newcommand{\eop}{{E^{\rm obs}_{\rm peak}}}
\newcommand{\eiso}{{E_{\rm iso}}}
\newcommand{\liso}{{L_{\rm iso}}}
\newcommand{\ep}{{E_{\rm peak}}}

\newcommand{\egamma}{{E_\gamma}}

\newlength{\GBCdigit}
\newcommand{\GBC}{\hspace*{\GBCdigit}}
\newlength{\GBCminus}
\newcommand{\Gbc}{\hspace*{\GBCminus}}

\shorttitle{Discovery of the short $\gamma$-ray burst GRB 050709}
\shortauthors{Villasenor et al.}

\begin{document}

\title{\bf Discovery of the short $\gamma$-ray burst GRB 050709}

\author{\rm
J.~S.~Villasenor$^{1}$,
D.~Q.~Lamb$^{2}$,
G.~R.~Ricker$^{1}$,
J.-L.~Atteia$^{3}$,
N.~Kawai$^{4}$,
N.~Butler$^{1}$,
Y. Nakagawa$^{5}$,
J.~G.~Jernigan$^{6}$,
M.~Boer$^{7}$,
G.~B.~Crew$^{1}$,
T.~Q.~Donaghy$^{2}$,
J.~Doty$^{8}$,
E.~E.~Fenimore$^{9}$,
M.~Galassi$^{9}$,
C.~Graziani$^{2}$,
K.~Hurley$^{6}$,
A.~Levine$^{1}$,
F.~Martel$^{10}$,
M.~Matsuoka$^{11}$,
J.-F.~Olive$^{7}$,
G.~Prigozhin$^{1}$,
T.~Sakamoto$^{12}$,
Y.~Shirasaki$^{13}$,
M.~Suzuki$^{14}$,
T.~Tamagawa$^{5}$,
R.~Vanderspek$^{1}$,
S.~E.~Woosley$^{15}$,
A.~Yoshida$^{6}$,
J.~Braga$^{16}$,
R.~Manchanda$^{17}$,
G.~Pizzichini$^{18}$,
K.~Takagishi$^{19}$,
~and~M.~Yamauchi$^{19}$,
}

\noindent\hangindent=6pt\hangafter=1
\hskip -24pt$^{1}${\small MIT Kavli Institute, Massachusetts Institute of
Technology, 70 Vassar Street, Cambridge, MA, 02139, USA.}

\noindent\hangindent=6pt\hangafter=1
$^{2}${Department of Astronomy and Astrophysics, University
of Chicago,\small  5640 South Ellis Avenue, Chicago, IL 60637, USA.}

\noindent\hangindent=6pt\hangafter=1
$^{3}${\small Laboratoire d'Astrophysique, Observatoire
Midi-Pyr\'{e}n\'{e}es, 14 Ave. E. Belin, 31400 Toulouse, France.}

\noindent\hangindent=6pt\hangafter=1
$^{4}${\small Department of Physics, Tokyo Institute of Technology, 
2-12-1 Ookayama, Meguro-ku, Tokyo 152-8551, Japan.}

\noindent\hangindent=6pt\hangafter=1
$^{5}${\small Department of Physics and Mathematics, Aoyama
Gakuin University, Fuchinobe 5-10-1, Sagamihara, Kanagawa 229-8558,
Japan.}

\noindent\hangindent=6pt\hangafter=1
$^{6}${\small University of California at Berkeley,
Space Sciences Laboratory, Berkeley, CA, 94720-7450, USA.}

\noindent\hangindent=6pt\hangafter=1
$^{7}${\small Centre d'Etude Spatiale des Rayonnements,
Observatoire Midi-Pyr\'{e}n\'{e}es,
9 Ave. de Colonel Roche, 31028 Toulouse Cedex 4, France.}

\noindent\hangindent=6pt\hangafter=1
$^{8}${\small Noqsi Aerospace, Ltd., 2822 South Nova Road,
Pine, CO 80470, USA.}

\noindent\hangindent=6pt\hangafter=1
$^{9}${\small Los Alamos National Laboratory, P.O. Box 1663, Los 
Alamos, NM, 87545, USA.}

\noindent\hangindent=6pt\hangafter=1
$^{10}${\small Espace Inc., 30 Lynn Avenue, Hull, MA 02045, USA.}

\noindent\hangindent=6pt\hangafter=1
$^{11}${\small Tsukuba Space Center, Japan Aerospace Exploration
Agency, Tsukuba, Ibaraki, 305-8505, Japan.}

\noindent\hangindent=6pt\hangafter=1
$^{12}${\small NASA Goddard Space Flight Center, Greenbelt, MD,
20771, USA.}

\noindent\hangindent=6pt\hangafter=1
$^{13}${\small National Astronomical Observatory, Osawa 2-21-1,
Mitaka,  Tokyo 181-8588 Japan.}

\noindent\hangindent=6pt\hangafter=1
$^{14}${\small RIKEN (Institute of Physical and Chemical Research),
2-1 Hirosawa, Wako, Saitama 351-0198, Japan.}

\noindent\hangindent=6pt\hangafter=1
$^{15}${\small Department of Astronomy and Astrophysics, University 
of California at Santa Cruz, 477 Clark Kerr Hall, Santa Cruz, CA
95064, USA.}

\noindent\hangindent=6pt\hangafter=1
$^{16}${\small Instituto Nacional de Pesquisas Espaciais, Avenida
Dos Astronautas 1758, S\~ao Jos\'e dos Campos 12227-010, Brazil.}

\noindent\hangindent=6pt\hangafter=1
$^{17}${\small Department of Astronomy and Astrophysics, Tata 
Institute of Fundamental Research, Homi Bhabha Road, Mumbai, 400 005, 
India.}

\noindent\hangindent=6pt\hangafter=1
$^{18}${\small INAF/IASF Bologna, Via Gobetti 101, 40129 Bologna, Italy.}

\noindent\hangindent=6pt\hangafter=1
$^{19}${\small Faculty of Engineering, Miyazaki University, Gakuen
Kibanadai Nishi, Miyazaki 889-2192, Japan.}

\clearpage

{\bf
Gamma-Ray Bursts (GRBs) fall into two classes: short-hard and 
long-soft bursts$^{1,2,3}$. The latter are now known to have X-ray$^{4}$ 
and optical afterglows$^{5}$, to occur at cosmological distances$^{6}$ in
star-forming galaxies$^{7}$, and to be associated with the explosion of
massive stars$^{8,9}$.  In contrast, the distance scale, the energy
scale, and the progenitors of short bursts have remained a mystery. 
Here we report the discovery of a short-hard burst whose accurate
localization has led to follow-up observations that have identified the
X-ray afterglow$^{10}$ and (for the first time) the optical
afterglow$^{10-11}$ of a short-hard burst.  These, in turn, have led to
identification of the host galaxy of the burst as a late-type galaxy at
$z = 0.16$,$^{10}$ showing that at least some short-hard bursts occur
at cosmological distances in the outskirts of  galaxies, and are likely
to be due to the merging of compact binaries.
}

On 9 July 2005, at 22:36:37 UT, the Soft X-Ray Camera (SXC), Wide-Field 
X-Ray Monitor (WXM) and French Gamma Telescope (FREGATE) instruments on 
board the High Energy Transient Explorer 2 satellite (hereafter HETE$^{12}$) 
detected GRB 050709, a short-hard pulse followed by a long-soft bump 
from the same location.  Figure 1 shows the WXM and SXC localizations 
for the burst, and the location of the X-ray and optical afterglows.  
Figures 2 and 3 show the time history of the entire burst and of the 
short-hard pulse in several energy bands.

Figure 4 compares the best-fit spectral model and the spectral data for
the short-hard pulse.  Table \ref{table:spectral_parameters} gives the
best-fit spectral parameters for different time intervals during the
burst.  The spectrum of the short pulse is hard and that of the intense
first peak of the pulse (corresponding to the first 0.2 s of the burst) 
is even harder.  

The duration and the peak energy $\eop$ of the spectrum of the
short-hard pulse are consistent with those of short-hard
GRBs.$^{13,14}$  We note that its duration is much shorter and its
spectrum is much harder than these were for GRB 020531, the other
short-hard burst localized by HETE.$^{15}$  We also note that the time
history and the spectral properties of GRB 050709 are similar to those
of several Burst and Transient Source Experiment (BATSE) bursts,
including GRBs 921022, 990516, and 990712$^{16,17}$ (see Table 1). 
Table \ref{table:emission_properties} gives the emission properties of
the burst.  The gamma-ray to X-ray fluence ratio of the short-hard
pulse is 3.1, which is also consistent with those of BATSE short-hard
bursts.

The isotropic-equivalent energy $\eiso$ of the short-hard pulse is a
factor $\sim$ 1000 smaller than is typical of long GRBs and implies
that the energy $\egamma$ radiated by the short-hard pulse in $\gamma$
rays is at least 40 times less than is typical for long GRBs.$^{18}$ 
The very small value of $\eiso$ also places the short-hard pulse off of
the $\eiso-\ep$ relation found by Refs. 19 and 20 for long GRBs by a
factor of $\sim$ 1000 in $\eiso$.  The luminosity $\liso$ of the
short-hard pulse is also very small, and places it off of the
$\liso-\ep$ correlation found by Refs. 20 and 21 for long GRBs by a
factor of $\sim$ 100 in $\liso$.  These three results strengthen the
conclusion that GRB 050709 is not a long-soft burst.

There is evidence that many short-hard bursts observed by the BATSE
and Konus instruments exhibit a long-soft bump following the initial 
hard pulse that is similar to that seen by HETE in GRB 050709.$^{22-24}$  
We have already mentioned the similarity of the time history and 
spectral properties of GRB 050709 and those of GRBs 921022, 990516, 
and 990712.  Refs. 22 and 23 reported evidence that BATSE and Konus 
short-hard bursts, respectively, are followed by a 30-200 s period 
of long-soft emission having a spectrum that is consistent with 
that of the long-soft bump in GRB 050709, while Ref. 24 reported 
evidence that BATSE short-hard bursts show an excess of soft 
emission from 20 s to 100-300 s after the burst.  

The fluence of the long-soft bump is much greater than that of the
short-hard pulse, unlike what is seen in SGR giant flares.$^{25,26}$ 
In addition, we have analyzed the time history of the long-soft bump
and find no evidence for brightness oscillations of the kind that
characterize the long-soft bump of SGR giant flares$^{25,26}$ (see
Figure 2); however, we can place only a weak constraint on the
amplitude of any such oscillations because the signal-to-noise ratio of
the light curve of the long-soft bump is low.  We have also searched
for any evidence of stochastic variability of the long-soft bump and
find none (also see Figure 2).  Finally, we have searched for evidence
of spectral evolution during the long-soft bump and find none (see
Table 1).

The most natural interpretation of the long-soft bump is that it is 
the beginning of the afterglow.  Its time history and spectrum are
consistent with those expected for an afterglow, as is the lack of any
time variability or spectral evolution.  The ratio of the fluence in
the short-hard pulse to that in the short-hard pulse plus the long-soft
bump implies a radiative efficiency of $<$ 25\% for the prompt phase. 
If the peak of the long-soft bump at $\approx 100$ s corresponds to the
time at which the fireball decelerates, a consistent solution exists in
which $z = 0.16$, the GRB jet has an isotropic-equivalent kinetic
energy $E_{\rm KE} \sim 5 \times 10^{49}$ erg, a relativistic bulk
$\Gamma \approx 100$, and expands into a low-density medium having a
number density $n \sim 10^{-2}$ cm$^{-3}$ (see Figure 2).

The accurate location of GRB 050509b by the BAT and XRT on board Swift
led for the first time to the identification of the X-ray afterglow of
a short GRB.$^{27}$  However, the burst occurred in the direction of two 
merging clusters of galaxies, making it impossible to determine the host
galaxy of the burst and thus the redshift of the burst ($>$ 20 galaxies
lie within the XRT error circle for the X-ray afterglow of the burst),
let alone the location of the burst within the host galaxy.

The accurate location of GRB 050709 by the WXM and SXC on board HETE
has led to the identification of the X-ray afterglow,$^{10}$ and for
the first time, the identification in ground-based$^{10-11}$ and
HST$^{11}$ images of the optical afterglow of a short-hard burst. 
These have led to the first secure identification of a host galaxy:  a
late-type spiral galaxy lying at a redshift $z = 0.16$.$^{11}$  The
X-ray and optical afterglows lie at a projected distance of $\approx 3$
kpc from the center of the host galaxy and are therefore not coincident
with the brightest optical emission from the host galaxy (in contrast
to long bursts$^{28}$).  

These results constrain the nature of the central engine for GRB
050709, and by implication all short-hard bursts.  The absence of any
large-amplitude oscillations with a period in the range 1-10 s in the
long-soft bump and the offset of GRB 050709 from the center of the host
galaxy argue against an association between this short-hard burst and
SGR giant flares.$^{25,26}$  Models based upon core collapse in massive
stars$^{9}$ explain the association of some long-soft bursts with
supernovae.  However, given the time it takes for the GRB-producing jet
to emerge from a collapsing massive star, it is difficult for such
models to produce bursts shorter than a few seconds.  Merging neutron
stars, on the other hand, can produce very short bursts.$^{29}$  Given
the smaller mass of the accretion disk that forms, it is not
unreasonable to expect a lower average $\eiso$ for short-hard bursts,
and hence a smaller average redshift.  Moreover, since binary neutron
stars are imparted with a ``kick'' at birth and travel large distances
before merging, one expects an offset of order several kpc between the
star-forming regions of the host galaxy and the burst.$^{30}$  GRB
050709 exhibits all of these properties.  The roughly 100 s lag between
the short-hard pulse of GRB 050709 and the peak of the much softer
afterglow is consistent with the low-density interstellar medium
expected in the vicinity of a merging compact binary.$^{22}$  If
short-hard GRBs are due to merging neutron stars, they produce powerful
bursts of gravitational radiation that should be detectable by the 
second-generation Laser Interferometry Gravitational-Wave Observatory.

The HETE localization of the short-hard burst GRB 050709 has led to
follow-up observations that have identified the X-ray afterglow and
(for the first time) the optical afterglow of a short, hard burst. 
These, in turn, have led to identification of the host galaxy of the
burst as a late-type galaxy at $z = 0.16$, showing that at least some
short-hard bursts occur at cosmological distances in the outskirts of 
galaxies, and are likely to be due to the merging of compact binaries.

\bigskip
\clearpage

\noindent\hangindent=20pt\hangafter=1
1. Hurley, K. in {\it Gamma-Ray Bursts} (ed. W. Paciesas \& G. Fishman)
    3 (AIP, New York, 1992)

\noindent\hangindent=20pt\hangafter=1
2. Lamb, D. Q., Graziani, C. \& Smith, I. Evidence for two distinct
	morphological classes of gamma-ray bursts from their short time 
	scale variability. \apjn {\bf 413}, L11-14 (1993)	

\noindent\hangindent=20pt\hangafter=1
3. Kouveliotou, C., et al. Identification of two classes of
	gamma-ray bursts. \apjn {\bf 413}, L101-104 (1993)

\noindent\hangindent=20pt\hangafter=1
4. Costa, E., et al. Discovery of an X-ray afterglow associated
	with the gamma-ray burst of 28 February 1997.  \nature {\bf
	387}, 783-784 (1997)

\noindent\hangindent=20pt\hangafter=1
5. van Paradijs, J. et al. Transient optical emission from the error
	box of the gamma-ray burst of 28 February 1997.  \nature {\bf
	386}, 686-688 (1997)

\noindent\hangindent=20pt\hangafter=1
6. Metzger, M R. et al. Spectral constraints on the redshift of the
	optical counterpart to the gamma-ray burst of 8 May 1997. 
	\nature {\bf 387}, 878-879 (1997)

\noindent\hangindent=20pt\hangafter=1
7. Castander, F. \& Lamb, D. Q. A Photometric Investigation of the 
	GRB 970228 Afterglow and the Associated Nebulosity. \apjn 
	{\bf 523}, 593-601 (1999)


\noindent\hangindent=20pt\hangafter=1
8. Woosley, S. E. Gamma-ray bursts from stellar mass accretion
	disks around black holes. \apjn {\bf 405}, 273l-277 (1993)

\noindent\hangindent=20pt\hangafter=1
9. Stanek, C., et al. Spectroscopic Discovery of the Supernova 2003dh
	Associated with GRB 030329. \apjn {\bf 591}, L17-20 (2003)





\noindent\hangindent=20pt\hangafter=1
10. Fox, D. B., et al. ... \nature ,  submitted


\noindent\hangindent=20pt\hangafter=1
11. Hjorth, J., et al. The optical afterglow of a short $\gamma$-ray
	burst. \nature, submitted



\noindent\hangindent=20pt\hangafter=1
12. Ricker, G. R., et al. in {\it Gamma-Ray Burst and Afterglow
    Astronomy 2001: A Workshop Celebrating the First Year of the HETE
    Mission} (ed. G. R. Ricker and R. K. Vanderspek) 3-16 (AIP Press,
    New York, 2003)



\noindent\hangindent=20pt\hangafter=1
13. Paciesas, W., et al. in {\it Gamma-Ray Bursts in the Afterglow
	Era} (ed. E. Costa, F. Frontera and J. Hjorth) 13-15 (Springer, 
	Berlin, 2000)

\noindent\hangindent=20pt\hangafter=1
14. Ghirlanda, G., Ghisellini, G. \& Celotti, A., The spectra of short
    gamma-ray bursts. \aan {\bf 422}, L55-58 (2004)

	
\noindent\hangindent=20pt\hangafter=1
15. Lamb, D. Q., et al. in {\it Gamma-Ray Bursts in the Afterglow Era}
    (ed. M. Feroci, F. Frontera, N. Masetti, and L. Piro) 94-97 (ASP, 
    San Francisco, 2004). 

\noindent\hangindent=20pt\hangafter=1
16. Paciesas, W., et al. The Fourth BATSE Gamma-Ray Burst Catalog
	(Revised). \apjsn {\bf 122}, 465-495 (2000)

\noindent\hangindent=20pt\hangafter=1
17. Preece, R., et al. The BATSE Gamma-Ray Burst Spectral Catalog. I.
    High Time Resolution Spectroscopy of Bright Bursts Using High
    Energy Resolution Data. \apjsn, {\bf 126}, 19-36 (2000)

\noindent\hangindent=20pt\hangafter=1
18. Frail, D. A., et al. Beaming in Gamma-Ray Bursts:  Evidence for a
	Standard Energy Reservoir. \apjn {\bf 562}, L55-58 (2001)

\noindent\hangindent=20pt\hangafter=1
19. Amati, L., et al. Intrinsic spectra and energetics of BeppoSAX
    Gamma-Ray Bursts with known redshifts. \aan {\bf 390}, 81-89 
    (2002)

\noindent\hangindent=20pt\hangafter=1
20. Lamb, D. Q., et al. Scientific Highlights of the HETE-2 Mission. 
   {\it New Astron. Rev.} {\bf 48}, 423-430 (2004)

\noindent\hangindent=20pt\hangafter=1
21. Yonetoku, D., et al. Gamma-Ray Burst Formation Rate Inferred from the
    Spectral Peak Energy-Peak Luminosity Relation. \apjn {\bf 609},
    935-951 (2004)

\noindent\hangindent =20pt\hangafter=1
22. Lazzati, D., Ramirez-Ruiz, E. \& Ghisellini, G. Possible detection
    of hard X-ray afterglows of short $\gamma$-ray bursts. \aan {\bf
    379}, L39-43 (2001)

\noindent\hangindent =20pt\hangafter=1
23. Frederiks, D. D., et al. in {\it Gamma-Ray Bursts in the Afterglow Era}
    (ed. M. Feroci, F. Frontera, N. Masetti, and L. Piro) 197-200 (ASP, 
    San Francisco, 2004). 

\noindent\hangindent=20pt\hangafter=1
24. Connaughton, V. BATSE Observations of Gamma-Ray Burst Tails. \apjn
    {\bf 567}, 1028-1036 (2002)

\noindent\hangindent=20pt\hangafter=1
25. Hurley, K., et al. An exceptionally bright flare from SGR
	1806-20 and the origin of short duration gamma-ray bursts.
	\nature {\bf 434}, 1098-1103 (2005)

\noindent\hangindent=20pt\hangafter=1
26. Palmer, D. M., et al. A giant gamma-ray flare from the magnetar
	SGR 1806-20. \nature {\bf 434}, 1107-1109 (2005)

\noindent\hangindent =20pt\hangafter=1
27. Gehrels, N., et al. The first short gamma-ray burst localized by
        Swift. \nature {\bf },  (2005)

\noindent\hangindent=20pt\hangafter=1
28. Bloom, J., et al. The Observed Offset Distribution of Gamma-Ray
    Bursts from Their Host Galaxies: A Robust Clue to the Nature of the
    Progenitors. \ajn {\bf 123}, 1111-1148 (2002)

\noindent\hangindent =20pt\hangafter=1
29. Eicher, D., Livio, M., Piran, T., \& Schramm, D. N. 
	Nucleosynthesis, neutrino bursts and $\gamma$-rays from coalescing
	neutron stars. \nature {\bf 340}, 126-128 (1989)






\noindent\hangindent =20pt\hangafter=1
30. Fryer, C.~L., Woosley, S.~E. \& Hartmann, D.~H. Formation Rates of
    Black Hole Accretion Disk Gamma-Ray Bursts. \apjn, {\bf 526}, 152-177 
    (1999)

\noindent\hangindent=20pt\hangafter=1
31. Reichart, D. E., et al. A Possible Cepheid-like Luminosity Estimator
	for the Long Gamma-Ray Bursts. \apjn {\bf 552}, 57 (2001)


\bigskip

\noindent{\bf Acknowledgements} 
This research was supported in the U.S. by NASA.

\noindent{\bf Author Information} The authors declare no competing financial
interests. Correspondence should be addressed to G.R.R.
(grr@space.mit.edu) or to J.S.V. (jsvilla@space.mit.edu).

\clearpage

\begin{deluxetable}{lcccc}
\tablecaption{Spectral Model Parameters for GRB 050709.
\label{tbl:spectrum}}
\tablewidth{0pt}
\tablehead{
\colhead{Parameter}
                    & \colhead{t = 0--0.20~s}
                    & \colhead{t = 0.20--0.50~s}
		    & \colhead{t = 0--0.50~s}
		    & \colhead{t = 20--180~s}
}
\startdata
Spectral Model:  & PLE & PLE & PLE & PL \\
\GBC {\small Photon index $\alpha$} &
 			$ \Gbc 0.53^{+0.12      }_{-0.13} $  &  
 			$ \Gbc 0.55^{+1.0       }_{-1.3}  $  &  
 			$ \Gbc 0.82^{+0.13      }_{-0.14} $  &  
 			$ \Gbc 1.98^{+0.18      }_{-0.15} $  \\ 
\GBC {\small Peak energy ($\eop$)}          &
 			$ \Gbc \GBC83.9^{+11\GBC\GBC}_{-8.3}  $ & 
 			$ \Gbc \GBC10.6^{+4.5\GBC\GBC}_{-3.5} $ & 
 			$ \Gbc \GBC86.5^{+16\GBC\GBC}_{-11}   $ & 
 			--- \\ 
\GBC {\small Normalization (at 15 keV)}             &
 			$ \Gbc 0.79^{+0.07\GBC    }_{-0.08}    $  & 
 			$ \Gbc 0.650^{+3.24\GBC   }_{-0.48}    $  & 
 			$ \Gbc 0.377^{+0.04\GBC   }_{-0.04}    $  & 
 			$ \Gbc 0.0075^{+0.0013\GBC   }_{-0.0013} $  \\
Chi-squared (DOF) & 439 (366) & 374 (366) & 467 (366) & 336 (367) \\ 
\enddata
\vskip -18pt
\tablecomments{The spectral models are power-law times exponential
(PLE) and power-law (PL).  Errors are for 90\% confidence.  The
normalization units are photons cm$^{-2}$ s$^{-1}$ keV$^{-1}$.  DOF is
the number of degrees of freedom in the fit.  We have also fit PL and
Band models to the first three time intervals and find $\chi^2$ (DOF)
values of 550 (367), 383 (367), and 538 (367) for the PL model and 439
(365), 374 (365), and 467 (365) for the Band model, demonstrating that
the data request the PLE model but not the Band model.  Similarly, we
have also fit a PLE model to the fourth time interval and find 
$\chi^2$ (DOF) values of 336 (366), demonstrating that the data request
the PL model but not the PLE model.  The large values of $\chi^2$ per
DOF that we find for the PLE model for the 0-0.2 s and 0-0.5 s time
intervals are due to the rapid spectral evolution during the short-hard
pulse.  We have calculated the hardness ratio $H$ = counts(5-10
keV)/counts(2-5 keV) for 10, 20, 30, and 40 s time intervals and find
no evidence of spectral evolution during the long-soft bump.  The time
history and spectral parameters of GRB 050709 are similar to those of
the BATSE bursts GRBs 921022, 990516, and 990712.$^{17,18}$  In
particular, the short-hard pulse of GRB 921022 had a duration $\approx$
256 ms and spectral parameters $\alpha=-1.1 \pm 0.3$, $\beta = -2.15
\pm 0.12$, and $\eop = 123 \pm 28$ keV, and the long-soft bump of that
burst had a PL spectrum with index $\alpha \approx$ -2;$^{17}$ while
the short-hard pulse of GRB 990712 had a duration $\approx$ 0.75 s and
spectral parameters $\alpha \approx -0.2$ and $E_0 \approx 600$ keV, 
and the long-soft bump of that burst had a PL spectrum with index
$\alpha = -1.9 \pm 0.6$.$^{23}$
}
\label{table:spectral_parameters}
\end{deluxetable}

\begin{deluxetable}{rccccc}
\tablecaption{Emission Properties of GRB 050709.
\label{tbl:emission}}
\tablewidth{0pt}
\tablehead{
\colhead{{\small Energy}} &
\colhead{{\small Peak Photon Flux}} &
\colhead{{\small Photon Fluence}} &
\colhead{{\small Peak Energy Flux}} & 
\colhead{{\small Energy Fluence}} \\
\colhead{{\small (keV)}} & \colhead{{\small (ph cm$^{-2}$ s$^{-1}$)}} & 
\colhead{{\small (ph cm$^{-2}$)}} &
\colhead{{\small ($10^{-8}$ erg cm$^{-2}$ s$^{-1}$)}} & 
\colhead{{\small ($10^{-8}$ erg cm$^{-2}$)}}
}
\startdata
{\small Short pulse:} & & & & \\
   2-10 & $29.0 \pm   5.2$   & $3.47 \pm   0.59$  & $24.9 \pm 3.9$ & $2.81 \pm   0.42$ \\
   2-25 & $53.6 \pm   6.1$   & $5.55 \pm   0.69$  & $88.7 \pm 7.7$ & $8.31 \pm   0.70$ \\
   2-30 & $58.1 \pm   6.9$   & $5.94 \pm   0.70$  & $111  \pm 8.6$ & $10.1 \pm   0.76$ \\
   7-30 & $37.1 \pm   2.8$   & $3.31 \pm   0.25$  & $96.8 \pm 6.8$ & $8.35 \pm   0.59$ \\
 30-400 & $34.1 \pm   2.7$   & $2.51 \pm   0.22$  & $400  \pm 46$  & $30.3 \pm   3.8$ \\
 50-100 & $13.9 \pm   1.1$   & $0.986 \pm  0.087$ & $156  \pm 13$  & $11.0 \pm   1.0$ \\
100-300 & $6.62 \pm   1.1$   & $0.515 \pm  0.092$ & $155  \pm 29$  & $12.4 \pm   2.5$ \\
  2-400 & $92.1 \pm   7.6$   & $8.43 \pm   0.752$ & $511  \pm 49$  & $40.3 \pm   4.1$ \\

{\small Long bump:}& & & & \\
 2-10 & $2.36 \pm 0.43$ & $107 \pm 18$ & $1.53 \pm 0.27$ & $69.1 \pm 10$ \\
 2-25 & $2.72 \pm 0.47$ & $123 \pm 18$ & $2.41 \pm 0.37$ & $109 \pm 14$ \\

\enddata
\vskip -18pt
\tablecomments{The quantities in this table are derived assuming the
best-fit PLE model for the spectrum of the short-hard pulse and the
best-fit PL model for the spectrum of the long-soft bump.  Errors are
90\% confidence level. The photon number and photon energy peak fluxes
for the short-hard pulse were evaluated over a 70 ms interval, 
corresponding to $T_{90}$ for the short-hard pulse; those for the 
long-soft bump are evaluated in a 1 s interval.  Using the
redshift $z = 0.16$ measured for the host galaxy,$^{11}$ the
isotropic-equivalent energy of the short-hard pulse in the 1-10,000 keV
energy band in the rest frame of the source is $\eiso =
(2.8^{+0.4}_{-0.2}) \times 10^{49}\ {\rm erg}$, taking $\Omega_M=0.3$,
$\Omega_\Lambda=0.7$, and $h=0.65$. Using a time interval 0.060 s in
the rest frame of the source (corresponding to a duration of 0.07 s in
the observer frame) and assuming the same cosmology and energy band,
the luminosity of the short-hard pulse in the 1-10,000 keV energy band
in the rest frame of the source is $\liso = (5.2 \pm 0.7) \times
10^{50}\ {\rm erg}\ {\rm s}^{-1}$. 
}
\label{table:emission_properties}
\end{deluxetable}

\clearpage 

\centerline{FIGURE CAPTIONS}

\noindent Fig. 1.--{\bf Sky map showing the HETE localization error
circles for GRB 050709 and the location of the X-ray and optical
afterglow.}  The WXM obtained a localization in flight.  However, the
spacecraft attitude-control system was not locked at the time of the
trigger, resulting in a drift of the satellite pointing direction, and 
real-time aspect was not available.  Consequently, the location was not
distributed in real time.  Ground analysis of the data from the optical
cameras provided reliable spacecraft aspect information, despite the
spacecraft drift rate.  A GCN Notice was sent out at 22:00:09 UT on 10
July 2005, after ground determination of the spacecraft aspect. 
Ground analysis of the WXM data produced a location with a 90\%
confidence region that is a circle centered at
R.A. = +23h 01m 44s ; Dec. = -38$^\circ$ 59$^{\arcmin}$ 52$^{\arcsec}$ (J2000)
%
with a radius of 14.5$^{\arcmin}$ (large circle labelled ``WXM-ground'').
Ground analysis of the SXC data  yielded a refined location with a 90\%
confidence region that is a circle centered at
R.A. = +23h 01m 30s ; Dec. = -38$^\circ$ 58$^{\arcmin}$ 33$^{\arcsec}$ (J2000)
%
with a radius of 1.34$^{\arcmin}$ (small circle labelled ``SXC-ground'').
The location of the X-ray and optical afterglow is labelled
``afterglow.''

\noindent Fig. 2.--{\bf Time history of GRB 050709.} From top to
bottom:  Time history observed by WXM in the 2-10 keV energy band  (a)
and in the 2-25 keV energy band (b); time history observed by FREGATE
in the 6-40 keV energy band (c) and in the 30-400 keV energy band (d). 
The event is a short-hard spike of duration $T_{90}$ = 220 $\pm$ 50 ms in
the 2-25 keV energy band and 70 $\pm$ 10 ms in the 30-400 keV energy
band, followed $\sim$ 25 seconds later by a long-soft bump of duration
$T_{90}$ = 130 $\pm$ 7 s in the 2-25 keV energy band -- where $T_{90}$ 
is the time interval containing 90\% of the photons.  We have performed 
an FFT on the time history of the long-soft bump for the time interval 5-175 s
after the trigger.  We find no evidence for any coherent brightness
oscillations in the period range 1-10 s, and derive 3$\sigma$ upper
limits on the amplitude of any such oscillations of 77\% and 89\% at
periods of 1 and 5 s, respectively.  We have also computed the
variability measure $V$,$^{27}$ for the long-soft bump, using smoothing
time scales of 10, 20, 30, and 40s.  We find $V = 0.005 \pm 0.013,
0.003 \pm 0.026, 0.007 \pm 0.031$, and $0.025 \pm 0.036$,
respectively.  We therefore find no evidence for a non-zero value of
$V$.  If the long-soft bump is the afterglow and its peak at $t_{\rm
peak} \approx 100$ s corresponds to the time at which the fireball
decelerates, a consistent solution exists in which $z = 0.16$; the GRB
jet has an isotropic-equivalent kinetic energy $E_{\rm KE} = (1-\cos
\theta_{\rm jet}) \eiso^{\rm total}/\eta \sim 5 \times 10^{49}$ erg --
where $\eiso^{\rm total}= \eiso({\rm pulse}) + \eiso({\rm bump}) \sim 1
\times 10^{50}$ erg, $\theta_{\rm jet} \sim 0.3$ is the jet opening
angle,$^{18}$ and $\eta = 0.2$ is the  radiative efficiency; a
relativistic bulk $\Gamma \approx 100$; and expands into a low-density
medium having a number density $n \sim 10^{-2}$ cm$^{-3}$.$^{22}$

\noindent Fig. 3.--{\bf Time history of the short-hard pulse of GRB
050709.}  From top to bottom:  Time histories observed by the WXM in
2--10 kev energy band (a) and 10--25 keV band (b); and by FREGATE in 
6--30 keV energy band (c), 30--85 keV energy band (d), and 85--400 keV
energy band (e), plotted in 5 ms time bins.  The pulse has a duration
$T_{90}$ = 220 $\pm$ 50 ms in the 2-25 keV energy band and 70 $\pm$ 10 ms 
in the 30-400 keV energy band, and exhibits no detectable emission before
$T=0$ or after $T=400$ ms, confirming the  short, hard nature of the
pulse.

\noindent Fig. 4.--{\bf Comparison of the observed count spectrum and
the best-fit PLE model for the short-hard pulse of GRB 050709.}  Upper panel:
comparison of the counts in the WXM energy loss channels (lower
energies) and the FREGATE energy loss channels (higher energies) and
those predicted by the best-fit PLE model (smooth curves). 
Error bars are one sigma (i.e. 68\% confidence limits). Lower panel: 
residuals to the fit. Error bars are one sigma. The short-hard pulse 
exhibits emission at
all energies, confirming that its spectrum is hard.

\clearpage

\begin{figure}
\includegraphics[scale=1.8,clip=]{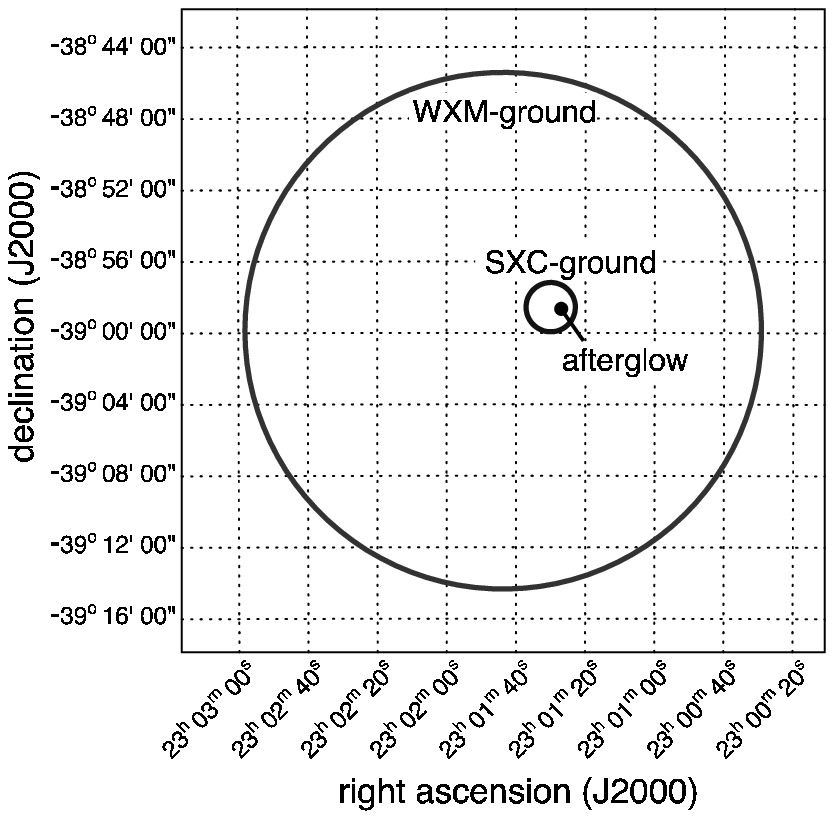}
\label{fig:skymap}
\end{figure}

\clearpage

\begin{figure}
\includegraphics[scale=0.85,clip=]{wxm_fre_thxg_lc.ps}
\label{fig:burst_time_histories}
\end{figure}

\clearpage

\begin{figure}
\includegraphics[scale=0.85,clip=]{wxm_fre_5chan.ps}
\label{fig:pulse_time_histories}
\end{figure}

\clearpage


\begin{figure}
\includegraphics[angle=0,scale=1.8,clip=]{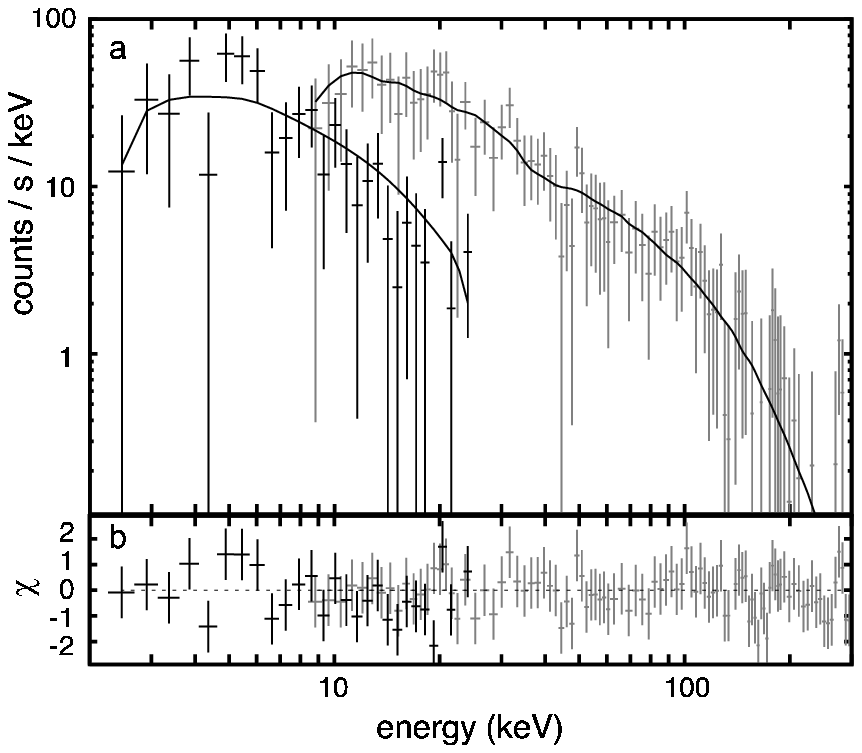}
\label{fig:spectra}
\end{figure}

\end{document}